\documentclass[prl,reprint,superscriptaddress,showpacs,showkeys,preprintnumbers,nofootinbib]{revtex4-1}
\usepackage{amsmath,amsfonts,amssymb,amsbsy}
\usepackage{mathrsfs,latexsym}
\usepackage{graphicx}
\usepackage{color,colortbl}
\definecolor{darkred}{rgb}{0.4,0.0,0.0}
\definecolor{darkgreen}{rgb}{0.0,0.4,0.0}
\definecolor{darkblue}{rgb}{0.0,0.0,0.4}
\usepackage[bookmarks,linktocpage,colorlinks,
    linkcolor = darkred,
    urlcolor  = darkblue,
    citecolor = darkgreen
    ]{hyperref}
\usepackage{macros_alpha}
\usepackage{dcolumn}
\usepackage{booktabs}
\newcolumntype{C}{>{$}c<{$}}
\AtBeginDocument{
\heavyrulewidth=.08em
\lightrulewidth=.05em
\cmidrulewidth=.03em
\belowrulesep=.65ex
\belowbottomsep=0pt
\aboverulesep=.4ex
\abovetopsep=0pt
\cmidrulesep=\doublerulesep
\cmidrulekern=.5em
\defaultaddspace=.5em
}
\usepackage{multirow}

\usepackage{defs}

\begin{document}

\newcommand{\LUV}{\mu_{\rm cut}}
\newcommand{\mupt}{\mu_{\rm PT}}
\newcommand{\muhadi}[1]{\mu_{\rm had #1}}
\newcommand{\muref}{\mu_{\rm phys}}

\newcommand{\full}{{(\nf)}}
\newcommand{\eff}{{(0)}}

\newcommand{\tq}{\full}
\newcommand{\tl}{\eff}

\newcommand{\Lameff}{\Lambda^\eff}
\newcommand{\betanl}{\beta_{\tl}}
\newcommand{\betanq}{\beta_{\tq}}

\newcommand{\gbarl}{\overline{g}^{\ell}}
\newcommand{\gbarf}{\overline{g}^\mathrm{f}}
\newcommand{\Lamf}{\Lambda_\msbar^\full}
\newcommand{\Laml}{\Lambda_\msbar^\eff}
\newcommand{\LamXfull}{\Lambda_X^\full}
\newcommand{\LamXeff}{\Lambda_X^\eff}
\newcommand{\LamX}{\Lambda_X}
\newcommand{\gbarX}{\overline{g}_X}
\newcommand{\gbarGF}{\overline{g}_\mathrm{GF}}
\newcommand{\gbarGFp}{\overline{g}_\mathrm{GF'}}

\newcommand{\FV}{\mathrm{FV}}
\newcommand{\gbarFV}{\overline{g}_{\FV}}
\newcommand{\FVp}{\mathrm{GF}}
\newcommand{\FVt}{\mathrm{GFT}}
\newcommand{\GFi}{\mathrm{GF\infty}}

\newcommand{\Plf}{P_{\ell,\rm f}}
\newcommand{\mudec}{\mu_\mathrm{dec}}

\newcommand{\GF}{gradient flow}


\preprint{WUB/19-05}
\preprint{DESY 19-224}
\title{
Non-perturbative renormalization by decoupling
}
\date{\today}
\collaboration{\href{https://www-zeuthen.desy.de/alpha/}{ALPHA collaboration}}

\author{Mattia~Dalla~Brida}
\affiliation{Dipartimento di Fisica, Universit\`a di Milano-Bicocca and \\
INFN, Sezione di Milano-Bicocca, Piazza della Scienza 3, 20126 Milano, Italy}
\author{Roman~H\"ollwieser} 
\affiliation{Department of Physics, Bergische Universit\"at Wuppertal, Gau{\ss}str. 20,
42119 Wuppertal, Germany}
\author{Francesco~Knechtli} 
\affiliation{Department of Physics, Bergische Universit\"at Wuppertal, Gau{\ss}str. 20,
42119 Wuppertal, Germany}
\author{Tomasz~Korzec} 
\affiliation{Department of Physics, Bergische Universit\"at Wuppertal, Gau{\ss}str. 20,
42119 Wuppertal, Germany}
\author{Alberto~Ramos} 
\affiliation{School of Mathematics and Hamilton Mathematics Institute, Trinity College Dublin, Dublin 2, Ireland}
\author{Rainer~Sommer}
\affiliation{John von Neumann Institute for Computing (NIC), DESY, Platanenallee~6, 15738~Zeuthen, Germany}
\affiliation{Institut~f\"ur~Physik, Humboldt-Universit\"at~zu~Berlin, Newtonstr.~15, 12489~Berlin, Germany}

\begin{abstract}
  We propose a new strategy for the determination of the QCD coupling. 
  It relies on a coupling computed in QCD with
   $\nf\geq3$ degenerate heavy quarks at a low energy scale 
  $\mudec$, together with a non-perturbative determination of the ratio
  $\Lambda/\mudec$ in the pure gauge theory. We explore this idea using a finite volume
  renormalization scheme for the case of $N_{\rm f}
  = 3$ QCD, demonstrating that a precise value of the strong coupling
  $\alpha_s$ can be obtained.  
  The idea is quite general and can be applied to solve other
  renormalization problems, using finite or 
  infinite volume intermediate renormalization schemes. 
\end{abstract}



\maketitle

\section{Introduction}
\label{sec:intro}

Currently the best estimates of $\alpha_s(m_Z)$ reach a precision
below $1\%$,
with lattice QCD providing the most precise determinations~\cite{Aoki:2019cca,Maltman:2008bx,Aoki:2009tf,McNeile:2010ji,Chakraborty:2014aca,Bazavov:2014soa,Nakayama:2016atf,Bruno:2017gxd}. 
The main challenge in a solid extraction of $\alpha_s$ by using lattice
QCD is the estimate of perturbative truncation uncertainties, other 
power corrections, and finite lattice spacing errors 
which are present in all extractions (see also~\cite{Brida:2016flw,DallaBrida:2018rfy}). 

A dedicated lattice QCD approach, known as step
scaling~\cite{Luscher:1991wu}, allows to connect an 
experimentally well-measured low-energy quantity
with the high energy regime of QCD
where perturbation theory can be safely applied, \emph{without making
  any assumptions on the physics at energy scales of a few GeV}.
It has recently been applied to three flavor QCD, yielding
 $\alpha_s(m_Z)$ with very high precision by means of a
non-perturbative running from scales of 0.2 GeV to 70
GeV~\cite{Bruno:2017gxd, DallaBrida:2016kgh, Brida:2016flw} 
and perturbation theory above. Although new
techniques~\cite{Luscher:2010iy, Fritzsch:2013je} have recently made
possible a significant improvement over older
computations~\cite{DellaMorte:2004bc,Aoki:2009tf,Tekin:2010mm}
a substantial further reduction of the overall error is challenging. 

In this paper we propose a new strategy for the computation of the
strong coupling. 
It is based on QCD with $\nf \geq 3$ quarks. 
We take the quarks to be degenerate, with an un-physically large mass, $M$.
They then decouple from the low-energy physics, which predicts our basic relation
\begin{eqnarray}
 \frac{\Lamf}{\mudec}\;P\left(\frac{M}\mudec \frac{\mudec}{\Lamf}\right) = 
 \frac{\Laml}{\Lambda_s^\eff}\, \varphi_s^\eff\left( \sqrt{u_\mathrm{M}}\right) + \rmO(M^{-2})\,, \nonumber \\[-2ex]
\label{eq:basic}
\end{eqnarray}
as we will explain in detail. Here $u_\mathrm{M}=\gbar^2_s(\mudec,M)$ is the value of the
coupling in a massive renormalization scheme at the scale
$\mudec$. The function
$\varphi^\eff_s( \gbar(\mudec))= \Lambda_s^\eff/\mudec$ relates
the same coupling and the renormalization scale $\mu=\mudec$ in
the zero-flavor theory and the function $P$ gives the ratio $\Laml/\Lamf$. As shown in 
\cite{Bruno:2014ufa,Athenodorou:2018wpk} $P$ is described
very precisely by (high order) perturbation theory.
The scale $\mudec$ has to be small compared to $M$ 
but is arbitrary otherwise. To make contact to physical units of MeV for the $\Lambda$-parameter,
 $\mudec$ has to be related to a physical mass-scale
such as $\muref=m_\mathrm{proton}$  (at physical quark masses). The use of intermediate
unphysical scales~\cite{Sommer:2014mea} is of course
possible.

In essence the above formula relates the $\nf$-flavor 
$\Lambda$ parameter to the pure gauge one by means 
of a massive coupling. 
Since perturbation theory is used only at the scale $M$,
it can be controlled by making $M$ sufficiently large. 

The main advantage of this approach is that the 
non-perturbative running 
of $\alpha_s$ from $\mudec$ to high 
energies is needed only in the
pure gauge theory, 
where high precision can be reached,
see~\cite{DallaBrida:2019wur}. It is connected
  to the three flavor theory  by a perturbative approximation for $P$, which is very
accurate already for masses around the charm mass, $M\approx
M_\mathrm{charm}$~\cite{Athenodorou:2018wpk}.   

Simulating heavy quarks on the lattice is a challenging multi-scale
problem, but defining the intermediate scheme, $s$, in a finite volume
allows us to reach large quark masses $M\approx M_\mathrm{bottom}$.  
 
\section{Decoupling of heavy quarks}
\label{sec:decoupling}
\newcommand{\ord}{\mathrm{O}}

On general grounds, the effect of heavy quarks is expected  to give
small corrections to low energy physics~\cite{Appelquist:1974tg}. 
Following~\cite{Weinberg:1980wa}, QCD with $N_{\rm f} $ heavy quarks
of renormalization group invariant (RGI) mass $M$ is well described by 
an effective theory at energy scales
$\mu\ll M$. 
By symmetry arguments, this theory is just the pure gauge theory~\cite{Bruno:2014ufa}. 
Thus, dimensionless low energy observables can be 
determined in the pure gauge 
theory -- up to small
corrections. In particular this holds true for renormalized 
couplings in massive renormalization schemes~\cite{thresh:BeWe},
\begin{eqnarray}
  \label{eq:mas_coupl_rel}
  \bar g_s^{(\nf)}(\mu,M) &=& \bar g_s^{(0)} (\mu) + \rmO\left(M^{-2}\right)\, . 
\end{eqnarray}
Here and below, $\rmO(M^{-k})$ stands for terms of
$\rmO((\mu/M)^k),\; \rmO((\Lambda/M)^k)$ where 
$k=1,2$, see below.
Parameterizing the fundamental ($\nf$-flavor) theory 
in a massless renormalization scheme such as $\msbar$, \eq{eq:mas_coupl_rel}
also relates the values of the fundamental and effective couplings
in the form~\cite{thresh:BeWe}
\begin{equation}
  \label{eq:pertdec}
  [\bar g^{(0)}_{\overline{\rm MS} }(m^\star)]^2 = [\bar g^{(N_{\rm f})}_{\overline{\rm MS} }(m^\star)]^2 \times
  C \left( \bar g^{(N_{\rm f})}_{\overline{\rm MS} }(m^\star)\right)\,.
\end{equation}
In the chosen $\overline{\rm MS} $ scheme, 
$C$ is perturbatively known including four loops
~\cite{Grozin:2011nk,Chetyrkin:2005ia,Schroder:2005hy, Kniehl:2006bg,Gerlach:2018hen} 
and with our particular choice of scale,\footnote{The running quark mass in scheme $s$ is denoted $\mbar_s$.} $m^\star = \mbar_{\overline{\rm MS} }(m^\star)$, 
the one-loop term vanishes,
\begin{equation}
  \label{eq:xi}
  C(\bar g ) = 1 + c_2(\nf) \bar g ^4 + c_3(\nf) \bar g ^6 + c_4(\nf) \bar g ^8 + \rmO(\bar g ^{10})\,.
\end{equation}

This relation between couplings provides a relation between the
$\Lambda$-parameters in the fundamental and effective
theories~\cite{Athenodorou:2018wpk}. 
Given the  $\beta$-function,
\begin{equation}
  \beta_s(\bar g_s ) =\mu\frac{{\rm d} \bar g_s (\mu)}{{\rm d} \mu} \,,
\end{equation}
in a (massless) scheme $s$, the $\Lambda$-parameters are defined by\footnote{In our notation, the perturbative 
expansion of the $\beta$-function is $\beta(x)=- x^3 (b_0 +b_1 x^2+\ldots)$.}
\begin{eqnarray}
\label{eq:LRGI1}
  \Lambda_s^{(N_{\rm f} )} &=& \mu \varphi_s^{(N_{\rm f} )}(\bar g_s (\mu))\,, \\ 
  \label{eq:varphi}
    \varphi_s^{(\nf)}(\gbar_s) &=& ( b_0 \gbar_s^2 )^{-b_1/(2b_0^2)} 
        \rme^{-1/(2b_0 \gbar_s^2)} \\ \nonumber
     && \times \exp\left\{-\int\limits_0^{\gbar_s} \rmd x\ 
        \left[\frac{1}{\beta_s(x)} 
             +\frac{1}{b_0x^3} - \frac{b_1}{b_0^2x} \right] \right\} \,.
\end{eqnarray}
Thus, 
\begin{equation}
  \label{eq:lambda_rat}
  \frac{\Lambda_\msbar^{(0)}}{\Lambda_\msbar^{(N_{\rm f} )}} 
  = P(M/\Lambda_\msbar^{(N_{\rm f} )})\,, 
\end{equation}
where
\begin{equation}
  \label{eq:Pdef}
  P(y) = \frac{\varphi_\msbar^{(0)}\left(g^\star(y)\, 
                       [C(g^\star(y))]^{1/2}\right)}
  {\varphi_\msbar^{(N_{\rm f} )}(g^\star(y))}\,, \quad y\equiv M/\Lambda_\msbar^{(N_{\rm f} )}\,.
\end{equation}
The function 
\begin{equation}
\label{eq:gstar}
  g^\star(M/\Lambda_\msbar^{(N_{\rm f} )}) = \bar g_\msbar ^{(N_{\rm f} )}(m^\star)
\end{equation}
is easily evaluated as explained in \cite{Athenodorou:2018wpk}. 
High precision is achieved by using the five-loop
$\beta$-function~\cite{vanRitbergen:1997va,Czakon:2004bu,Baikov:2016tgj,Luthe:2016ima,Herzog:2017ohr}.

Finally, 
the combination of eqs.(\ref{eq:LRGI1},\,\ref{eq:pertdec},\,\ref{eq:lambda_rat}) results in
\begin{eqnarray}
 \label{eq:basic2}
 \rho \,P(z/\rho) &=& 
 \frac{\Laml}{\Lambda_s^\eff}\, \varphi_s^\eff( \sqrt{u_\mathrm{M}}) + \ord(M^{-2})\,,
 \\
u_\mathrm{M} &=& \gbar_s^2(\mudec,M)\,,
\end{eqnarray}
written in terms of the dimensionless variables
\begin{equation}
 \\
 \rho = \frac{\Lamf}{\mudec}\,, \quad z= M/\mudec
    \,.
\end{equation}
The current perturbative uncertainty in 
$P(M/\Lambda)$ is of $\rmO(\gbar^{8}(m^\star))$. 
It vanishes 
together with the power corrections of order $M^{-2}$ 
as $M$ is taken large. 
This completes the explanation of \eq{eq:basic}.

When evaluating  the above quantities by lattice simulations, a
multitude of mass scales are relevant: 
\begin{itemize}
	\item $1/L$, the inverse linear box size,
	\item $\mpi$, the pion mass,
	\item $\muref\sim\mudec\sim m_\mathrm{proton}$, typical QCD mass scales,
	\item $M$, 
	\item $a^{-1}$, the inverse lattice spacing.  
\end{itemize}
Small finite size effects require
$1/L\ll\mpi$, accurate decoupling is given when
$M \gg \mudec$ and all scales have to be small
compared to $a^{-1}$. 
Such multi-scale problems are very challenging;
they inevitably require very large lattices~\cite{Athenodorou:2018wpk}.

\subsection{Ameliorating the multi-scale problem with a finite volume strategy}

The multi-scale nature of the problem can be made manageable by using a finite volume coupling $\gbar_s(\mu) =\gbarFV(\mu)$ with \cite{Luscher:1991wu} 
\begin{equation}
	\mu=1/L\,.
\end{equation} 
The crucial advantages are:
\begin{enumerate}
\item 
  There is no need for the volume to be large.
\item We can choose an intermediate value for the scale 
  $\mudec$. With  $\mudec\approx 800$~MeV  large quark masses $M\approx 6000$ MeV can be simulated. 
  Then the uncertainties in the perturbative 
  evaluation of $P$ are negligible and the power corrections $(\mudec/M)^k$ are expected to be small~\cite{Athenodorou:2018wpk}.
\item One is free to choose a coupling definition that has a known
  non-perturbative running
in pure gauge theory, e.g. a gradient flow coupling~\cite{DallaBrida:2016kgh}. 
\end{enumerate}
It remains that $aM$ has to be small at large $M/\mudec$. 

Most finite volume couplings used in practice are 
formulated with Schr\"odinger functional  (SF) boundary 
conditions on the
gauge and fermion fields~\cite{Luscher:1992an, Sint:1993un} (i.e. 
Dirichlet boundary conditions in Euclidean time at $x_0=0,T$, and
periodic boundary conditions with period $L$ in the spatial
directions).
In this situation, the decoupling effective Lagrangian \cite{Athenodorou:2018wpk} contains 
terms with dimension four at the boundaries, which are suppressed by just one
power of $M$. We have to generalize the $\rmO(M^{-2})$ corrections in
\eq{eq:basic2} to 
$\rmO(M^{-k})$ where $k=1$ if a boundary is present \cite{Sint:1995ch}.
Finite volume schemes that preserve the invariance under translations,
using either periodic~\cite{Fodor:2012td} or twisted~\cite{Ramos:2014kla}
boundary conditions, would show a faster decoupling with $k=2$.
 
\section{Testing the strategy}

\label{sec:test}
\newcommand{\decoupstrat}[1]{      
  \begin{picture}(160,40)(-10,0)
  \put(-40,0){$\muref$}
  \put(-15,0){\vector(1,0){10}}
  \cblu\put(0,0){$ \gbar_\FVp^{(3)}(\mudec)$}
  \cbla\put(25,10){\vector(1,1){15}}
  \cred\put(25,30){$\gbar_\FVt^{(3)}(\mudec,M) = \gbar_\FVt^{(0)}(\mudec)$} 
  \cbla\put(110,23){\vector(1,-1){15}}
  \cblu\put(130,0){$\gbar_\FVp^{(0)}(\mudec)$}
  \end{picture}
}

We now turn to a numerical demonstration of the 
idea for $\nf=3$.
Our  
discretisation employs non-perturbatively $\rmO(a)$ improved
Wilson fermions, 
the same action as the CLS initiative~\cite{Bruno:2014jqa}.
The bare (linearly divergent) quark mass is denoted $m_0$ and 
the pure gauge action has a prefactor  $\beta=6/g_0^2$.
When connecting observables at different quark masses
it is important to keep the lattice spacing constant
up to order $(aM)^2$. This requires setting
$g_0^2=\gtilde^2/(1+\bg(\gtilde)a\mq)$, where $a\mq=am_0-a\mcrit$
and $a\mcrit$ denotes the point of vanishing quark mass.
The bare improved coupling
$\gtilde$ is independent of the quark
mass~\cite{Luscher:1996sc,Sint:1995ch}. We use the one-loop
approximation to $\bg$.

\subsection{Choice of finite volume couplings}

Several renormalized couplings can be defined in the SF  using the
Gradient Flow~\cite{Fritzsch:2013je} (see~\cite{Ramos:2015dla} for a
review of the topic). Our particular choice is based on 
\begin{equation}
  \label{eq:Edensity}
  E_\mathrm{mag}(t,x) = \frac{1}{4}G_{ij}^a(t,x)G_{ij}^a(t,x)\,,\quad (t>0; i,j=1,2,3)\,,
\end{equation}
i.e. the spatial components of the field strength\footnote{Using only the magnetic components reduces the boundary ${\rm O}(a)$ effects~\cite{Fritzsch:2013je}.}
\begin{equation}
  G_{\mu\nu}(t,x) = \partial_\mu B_\nu - \partial_\nu B_\mu + [B_\mu,B_\nu] 
\end{equation}
of the flow field defined by
\begin{equation}
  \partial_tB_\mu(t,x) = D_\nu G_{\nu\mu}(t,x)\,, \qquad B_\mu(0,x) = A_\mu(x)\,.
\end{equation}
Composite operators formed from the smooth 
flow field $B_\mu $ 
are finite~\cite{Luscher:2011bx} and thus
\begin{equation}
\label{eq:gfvp}
  [\bar g^{(3)}_\FVp(\mu)]^2 = \mathcal N^{-1} t^2\langle E_\mathrm{mag}(t,x) \rangle\Big|_{M=0, T=L}^{ x_0=L/2, \mu=1/L, \sqrt{8t}=cL}\,,
\end{equation}
is a finite volume renormalized coupling.  
Very precise results are available for
$\gbar^{(3)}_\FVp$
in $N_{\rm f} =3$ QCD~\cite{DallaBrida:2016kgh}
and in the Yang-Mills theory~\cite{DallaBrida:2019wur}.
The constant $\mathcal N$ is analytically
known~\cite{Fritzsch:2013je}, we take $c=0.3$ and project to 
zero topology~\cite{Fritzsch:2013yxa}; thus the coupling
is exactly the one denoted $\gbarGF$ in~\cite{DallaBrida:2016kgh}. 
However, 
it is advantageous to apply decoupling to a slightly different coupling,
\begin{equation}
  \label{eq:gmassive}
  [\bar g_{\FVt}^{(3)}(\mu,M)]^2 = \mathcal N'^{-1} t^2\langle E_\mathrm{mag}(t,x) \rangle\Big|_{T=2L}^{  x_0=L, \mu=1/L, \sqrt{8t}=cL}\,,
\end{equation}
where $E$ is inserted a factor two further away from
the boundary and the $M^{-1}$ effects are substantially
reduced~\cite{inprep}. In contrast to large changes in the renormalization scale, changes of the scheme,
$\gbar^2_{\FVp} \leftrightarrow \gbar^2_{\FVt} $ are 
easily accomplished numerically; they do not 
contribute significantly to the numerical effort or the
overall error. After choosing a precise value for $\mudec$ by
  fixing the value of $\bar g^{(3)}_\FVp(\mudec)$, the use of the two
  schemes is schematically shown in the graph
\begin{center}
\decoupstrat{2}
\end{center}
and explained in detail in the following section.

\begin{figure*}
  \centering
  \includegraphics[width=0.8\textwidth]{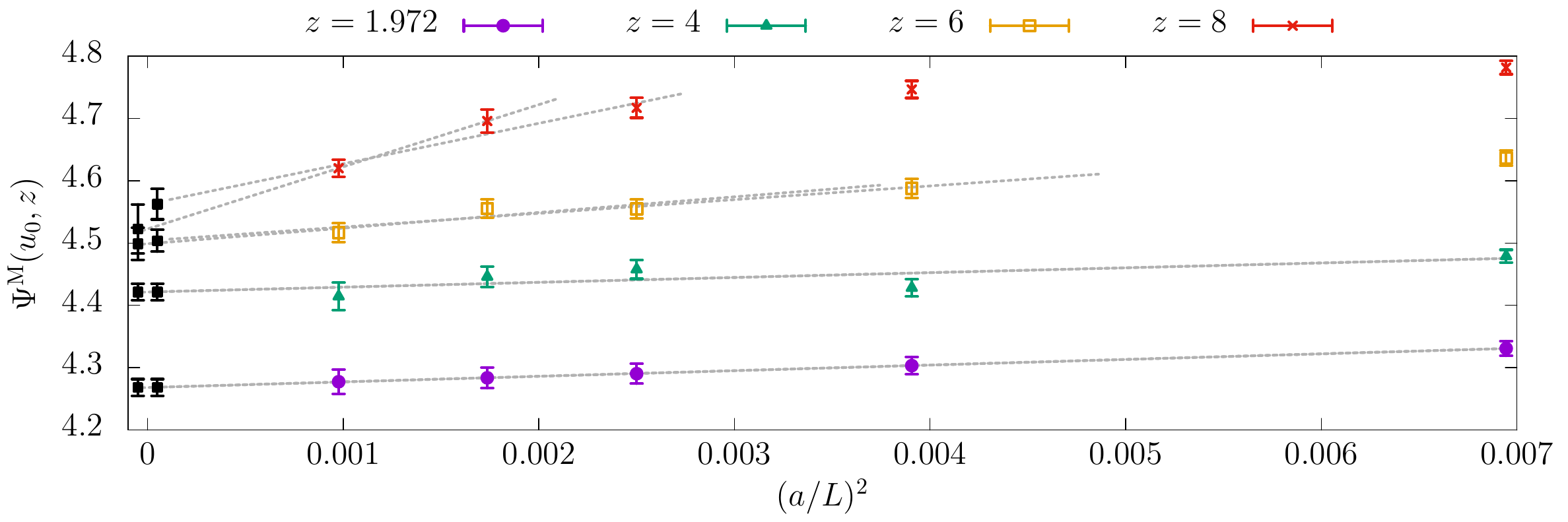} 
  \caption{Continuum extrapolation of the massive coupling $\Psi^\mathrm{M}(u_0,z)$.
    We apply two cuts $(aM)^2 < 1/8, 1/4$ in order to
    estimate the systematic uncertainty. 
  }
  \label{fig:cont}
\end{figure*}
\subsection{Numerical computation}
We fix a convenient value 
\begin{equation}
  \label{eq:mudec}
  [\bar g^{(3)}_\FVp(\mudec)]^2 = 3.95 \equiv u_0\,.
\end{equation}
With the non-perturbative 
$\beta$-function 
of~\cite{DallaBrida:2016kgh}
and the relation to the physical scale $\muref$ of 
\cite{Bruno:2017gxd,Bruno:2016plf}\footnote{The physical scale is set by
a linear combination of Pion and Kaon decay constants.} we deduce
\begin{equation}
  \mu_{\rm dec} = 789(15)\, {\rm MeV}\,.
\end{equation}
For this choice, the bare parameters, $\gtilde^2,\,am_0=a\mcrit(\gtilde^2)$ are known rather precisely for several resolutions $L/a$
~\cite{Fritzsch:2018yag}, see \Tab{tab:lcp}.

In order to switch to massive quarks of a given RGI mass,
$M=z/L$, we need to know $a\mq$ which is the solution of 
\begin{equation} 
  z 
  = \frac{L}{a}\, \frac{M}{\mbar(\mudec)}\,
    \zm(\gtilde,a/L)\cdot (1 + b_\mathrm{m}(\gtilde)\,a\mq)\, a\mq \,,
\end{equation}
where $\zm$ is the renormalization factor in the SF scheme employed in
~\cite{Campos:2018ahf} at scale $\mudec = 1/L$, the ratio
$\frac{M}{\mbar(\mudec)}=1.474(11)$ in the same scheme 
is derived from the results 
of~\cite{Campos:2018ahf}, and the term $b_\mathrm{m}\,a\mq$ 
removes the discretisation effects of $\rmO(aM)$. We have computed
$\zm,\, b_\mathrm{m}$, listed in  \tab{tab:lcp}, by dedicated simulations \cite{inprep}.

As indicated above, the switch to massive quarks is 
accompanied by the switch to $\gbar_\FVt$ in order to suppress linear $1/M$ terms: we evaluate
\begin{eqnarray}
  \Psi^\mathrm{M}(u_0,z) &=& \left[ \gbar_\FVt^{(3)}(\mudec,M) \right]^2_{[\gbar_\FVp^{(3)}(\mudec)]^2=u_0}\,,
  \\ && z=M/\mudec \nonumber\,.
\end{eqnarray}
Here, with bare mass $am_0$ 
set as  explained, the condition $[\gbar_\FVp^{(3)}(\mudec)]^2=u_0$ fixes $\gtilde^2$
to the values in \tab{tab:lcp}. 

\begin{table}
  \centering
  \begin{tabular}{lllllll}
    \toprule
    $L/a$ & $6/\gtilde^2$ & $\;a\mcrit(\gtilde^2)$ & $\;\;\;\gbar^2_{\FVp}$ & $\;\;\zm$ & $\;\;\;b_\mathrm{m}$\\
    \midrule
    12  & 4.3020 & $-0.3234(3)$ & 3.9533(59)  &1.691(7) & $-0.43(3)$ \\
    16  & 4.4662 & $-0.3129(2)$ & 3.9496(77)  &1.726(8) & $-0.50(3)$\\
    20  & 4.5997 & $-0.3043(3)$ & 3.9648(97) &1.741(10) & $-0.48(4)$ \\
    24  & 4.7141 & $-0.2969(1)$ &  3.959(50)  &1.770(11) & $-0.51(2)$\\
    32  &   4.90 & $-0.28543(4)$ &  3.949(11) &1.814(16) & $-0.63(5)$ \\
    \bottomrule
  \end{tabular} 
  \caption{At each $L/a$
  the bare coupling $\beta=6/\gtilde^2$ and the
  bare mass $am_0=a\mcrit$
  are fixed to have constant coupling, \eq{eq:mudec},
  and vanishing quark mass~\cite{Fritzsch:2018yag,matching:internal}, and $\zm,b_\mathrm{m}$
  are determined by simulations with different $a\mq$ at
  fixed $\gtilde$~\cite{inprep}.}
  \label{tab:lcp}
\end{table}
We repeat the exercise for $z=1.972, 4, 6, 8$, which correspond to $M\approx 1.6, 3.2, 4.7, 6.3$~GeV. 

It is left to perform continuum extrapolations of the function
$\Psi^\mathrm{M}(u_0,z)$, as illustrated in
\fig{fig:cont}. 
They become more challenging at large values of $z$. 
We explore the systematics by imposing
two mass cuts $(aM)^2<1/8,\, 1/4$ and find compatible results, with the
results with $(aM)^2<1/8$ having significantly larger errors,  at large values of $M$, where few points are left after the cut. 
We take the extrapolations using $(aM)^2<1/8$ as our best estimates of the continuum values of 
$\Psi^\mathrm{M}(u_0,z)$ (see second column of table~\ref{tab:results}). 

The precise non-perturbative $\beta$-function $\beta_\FVp^\eff$ of Ref.~\cite{DallaBrida:2019wur} 
determines $\varphi_\FVp^\eff(\gbar_\FVp)$ in the relevant range
of $\gbar_\FVp^2\gtrsim4$. We connect to it from the scheme $\FVt$ 
by extra simulations, which evaluate 
$\gbar_\FVt^{(0)}(\mu)$ at the same parameters $g_0^2,L/a$
where  
$\gbar_\FVp^{(0)}(\mu)$ is known. After continuum extrapolation 
of those data with $L/a=12,16,20,24$ we find for $3.8\leq [\gbar_\FVt^{(0)}]^{2}\leq 5.8$~\cite{inprep}
\begin{equation} 
	[\gbar_\FVp^{(0)}]^{-2}  - [\gbar_\FVt^{(0)}]^{-2} = p_0 +p_1 [\gbar_\FVt^{(0)}]^{2}
	+p_2 [\gbar_\FVt^{(0)}]^{4} \pm 7\times 10^{-4}\,, 
	\nonumber
\end{equation}
with $(p_0,p_1,p_2)=(2.886,-0.510,0.056)\times 10^{-2}$.
For each of the values $\Psi^\mathrm{M}$ in \tab{tab:results}
we obtain $u_\mathrm{M}=[\gbar_\FVp^{(0)}]^2$ from 
$[\gbar_\FVt^{(0)}]^2=\Psi^\mathrm{M}$, insert
into \eq{eq:basic2} (with scheme $s=\FVp$)
and solve (numerically) for $\rho$. The table includes $\Lambda_\msbar^{(3)}$ as well as
the influence of the last known term of
the series \eq{eq:xi} which demonstrates that perturbative
uncertainties are negligible.

At present we have used a relatively
modest amount of computer time. 
Our largest lattice is just $64\times 32^3$. A significant
improvement, simulating lattice spacings twice finer, is
possible with current computing power.
\begin{table}
  \centering
  \small
\begin{tabular}{llllcc}
   \toprule
   $z$ & $\Psi^\mathrm{M}$ &
       $\Lambda^{(0)}_{\overline{\rm MS} }/{\mu_{\rm  dec}}$
   & $\frac{1}{ P(M/\Lambda)}$ & $\Lambda^{(3)}_{\overline{\rm MS} }$ [MeV] & $\Delta_4$ [MeV]\\ 
   \midrule
   1.972& 4.268(13)&      0.689(11)&0.8000(48)&  434(12) & 2.0\\
   4.0&   4.421(13)&      0.725(11)&0.6865(28)&  393(11) & 0.7\\
   6.0&   4.499(26)&      0.743(13)&0.6283(26)&  368(10) & 0.4\\
   8.0&   4.523(40)&      0.749(14)&0.5889(27)&  348(11) & 0.3\\
   $\infty$ & \multicolumn{3}{c}{FLAG19 (lattice) \cite{Aoki:2019cca}}&   343(12) & \\
   \bottomrule
 \end{tabular}
  \caption{Results for the massive coupling $\Psi^\mathrm{M}(u_0,z)$
    at different values of $M$ and fixed $\mu_{\rm dec} = 789(15)$
    MeV. The perturbative factor $P(M/\Lambda)$ is determined 
  with five-loop running and including $c_{l\leq4}$ in \eq{eq:xi}. 
  $\Delta_4$ shows the effect of $c_4$ in $\Lambda_\msbar^{(3)}$. The effect of $c_3$ is larger by a factor 1.5 (for $z=1.972$) to 3
  (for $z=8$).}
  \label{tab:results}
\end{table}
\begin{figure*}
  \centering
  \includegraphics[width=0.85\textwidth]{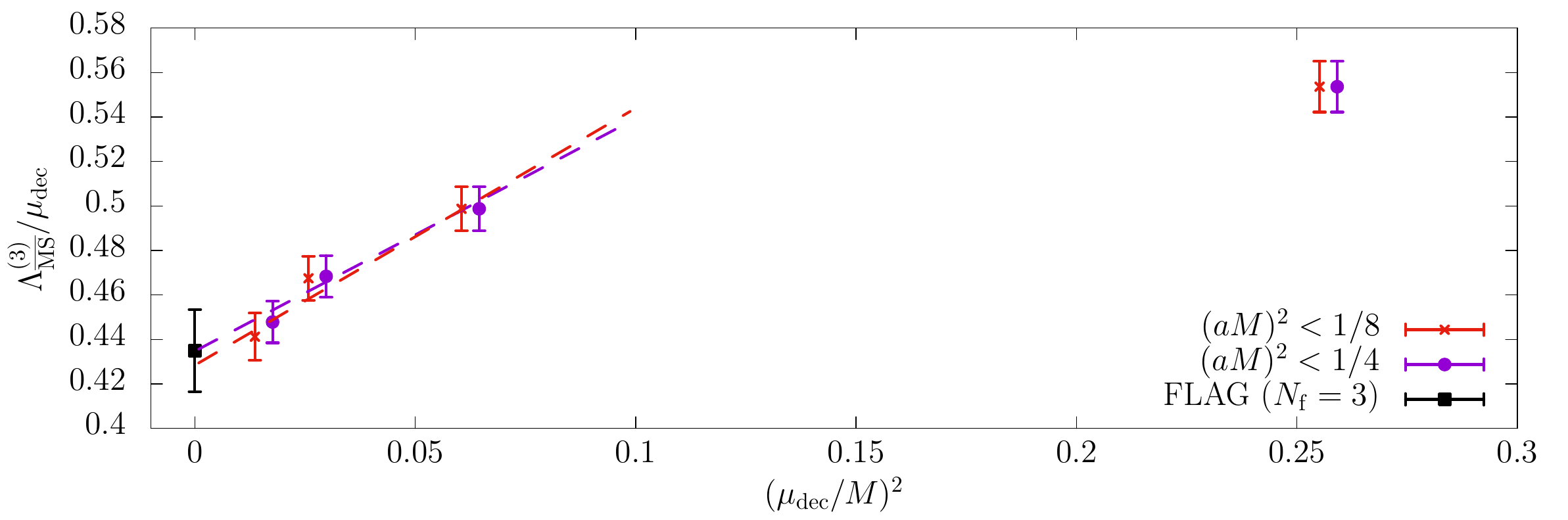} 
  \caption{Values for $\rho$ determined
    from the decoupling relation. 
    As $z=M/\mu_{\rm dec}$ gets larger, the
    approximations for $\rho = \Lambda_\msbar/\mudec$ approach the FLAG
    result for $\Lambda^{(3)}_\msbar$ in units of
    $\mudec=789(15)~\MeV$~\cite{Aoki:2019cca}.   
    The dashed lines illustrate possible extrapolations  $M\to\infty$ (cf. 
    Eq.~(\ref{eq:basic})). A significant part of the errors at finite
    $M$ is due to the Yang-Mills theory and may be reduced further.    
  \label{fig:final}}
\end{figure*}

\subsection{Results}

According to Eq.~(\ref{eq:basic}),
the values obtained for
$\Lambda^{(3)}_{\overline{\rm MS} }$  
approach the true non-perturbative value
as $M\to \infty$. We demonstrate this property
in the plot of $\rho$, Fig.~\ref{fig:final}. While we see 
 power corrections, these are small and the point with $M\approx 6$~GeV is in agreement with the known number from 
 \cite{Bruno:2017gxd} as well as with the FLAG average~\cite{Aoki:2019cca}. Rough extrapolations 
 to the limit $M\to\infty$ 
 seem to make the
 agreement even better. This limit should be
 studied with even higher precision in the future.

\section{Conclusions}
\label{sec:conclusions}

In this letter we propose a new strategy to determine the strong coupling. 
It requires the determination of a renormalized 
coupling in an
unphysical setup with degenerate massive quarks at some low energy scale. 
The second ingredient is the determination of the $\Lambda$-parameter 
in units of the low energy scale \emph{in the pure gauge theory} 
defined in terms of the same coupling. 
As we have shown, there is a clear advantage: the essential 
part of the multi-scale problem (i.e. 
the determination of the $\Lambda$-parameter) is done without fermions.
The remaining problem, namely the limit of large $M$ can be reached
by two observations. First it is known that with a mass $M$ of a few GeV,
the perturbative prediction for $P$ is {\em very} accurate
\cite{Herren:2017osy,Athenodorou:2018wpk}. Second we presented a finite 
volume strategy which allows to reach masses of several GeV and presented
clear evidence that the remaining power corrections are small. 
The result is in good agreement with the more standard step scaling
approach, but promises a higher precision.  
We note again that we only invested a rather modest numerical
effort. The limits $a\to0$ at fixed $M$ and $M\to\infty$ can be much
improved. 
With more work our new strategy will lead to
a substantial reduction of the uncertainty in $\alpha_s$. 
As mentioned, other definitions of the finite volume coupling with
other boundary conditions may be chosen.  

There is even a more direct approach, simply using $t_0$~\cite{Luscher:2010iy},
\begin{eqnarray}
 [\Lambda_\msbar\sqrt{t_0(M)}]^{(\nf)}\;P\left(\frac{M}{\Lamf}\right) = [\Lambda_\msbar\sqrt{t_0}]^{(0)}
     + \rmO(M^{-2})\,,
     \nonumber
\end{eqnarray}
or a different low energy scale
(see also~\cite{Bruno:2014ufa,Athenodorou:2018wpk}).
This only requires to determine such a gluonic
scale with at least three degenerate
massive quarks. 
Together with a determination of the pure gauge $\Lambda$-parameter in
units of the same scale, this simple strategy could possibly provide a
precise 
determination of the strong coupling. Controlling discretization
errors, power corrections and perturbative corrections at the same 
time will require compromises but is worth exploring. 

The idea presented here can easily be extended to other RGI
quantities. A clear case is the determination of quark masses. 
On the other hand,  four fermion operators require an investigation
from the start, exploring both perturbative and power corrections. 

\vskip1em
\begin{acknowledgments}
\emph{Acknowledgments---}
We are grateful to our colleagues in the ALPHA-collaboration for discussions and the sharing of 
code as well as intermediate un-published results. In particular we thank Stefan 
Sint for many useful discussions and P.~Fritzsch, J. 
Heitger, S. Kuberski for preliminary results of the HQET project \cite{Fritzsch:2018yag}. 
We thank Matthias Steinhauser for pointing out 
that $c_4$ is known \cite{Gerlach:2018hen}. 
RH was supported by the Deutsche Forschungsgemeinschaft in the SFB/TRR55.
AR and RS acknowledge funding by the H2020 program in the  {\em Europlex} training
 network, grant agreement No. 813942.
Generous computing resources were supplied 
by the North-German Supercomputing Alliance (HLRN, project bep00072) and by the John von
Neumann Institute for Computing (NIC) at DESY, Zeuthen.

\end{acknowledgments}

\bibliography{lambda-dec}

\end{document}